\documentclass[aps,pra,superscriptaddress,twocolumn,amssymb,amsmath,10pt]{revtex4-2}
\usepackage{amsmath}
\usepackage{amssymb}
\usepackage{latexsym}
\usepackage{revsymb}
\usepackage{leftindex}
\usepackage{enumitem}

\usepackage[colorlinks]{hyperref}

\hypersetup{
citecolor=magenta
}

\newcommand{\ket}[1]{\left | #1 \right \rangle}
\newcommand{\bra}[1]{\left \langle #1 \right |}
\newcommand{\braket}[2]{\left \langle #1 | #2 \right \rangle}
\newcommand{\braL}[2]{\leftindex_{#1}{\bra{#2}}}

\newcommand{\ketNoResize}[1]{| #1 \rangle}

\newcommand{\id}{\mbox{\bf 1}}

\newcommand{\beq}{\begin{equation}}
\newcommand{\eeq}{\end{equation}}
\newcommand{\ra}{\rangle}

\newcommand{\hU}{{\hat U}}
\newcommand{\hL}{{\hat L}}
\newcommand{\htheta}{{\hat \theta}}
\newcommand{\hp}{{\hat p}}
\newcommand{\hx}{{\hat x}}

\begin{document}
\title{Networks of quantum reference frames and the nature of conserved quantities}

\author{Daniel Collins}
\affiliation{H. H. Wills Physics Laboratory, University of Bristol, Tyndall Avenue, Bristol BS8 1TL, UK}

\author{Carolina Moreira Ferrera}
\affiliation{H. H. Wills Physics Laboratory, University of Bristol, Tyndall Avenue, Bristol BS8 1TL, UK}

\author{Ismael L. Paiva}
\affiliation{H. H. Wills Physics Laboratory, University of Bristol, Tyndall Avenue, Bristol BS8 1TL, UK}
\affiliation{Departamento de F\'isica, Centro de Ci\^encias Exatas e da Natureza,
Universidade Federal de Pernambuco, Recife, Pernambuco 50740-540, Brazil}

\author{Sandu Popescu}
\affiliation{H. H. Wills Physics Laboratory, University of Bristol, Tyndall Avenue, Bristol BS8 1TL, UK}


\date{March 26, 2026}

\begin{abstract}

We show that networks of quantum frames of reference, in which one frame may be used to produce multiple other frames that in their turn prepare systems which may interact with one another, have counterintuitive properties that make following the exchange of conserved quantities very subtle, and raise questions about the very nature of conserved quantities. In addition, we present an alternative approach to analysing quantum reference frames that we believe will be useful in discussions related to quantum frames of reference.

\end{abstract}

\maketitle

\section{Introduction}

In this paper we focus on the interplay between two important concepts in physics: conservation laws and reference frames. Specifically, we consider the recently discovered deeper relation that exists between these two concepts~\cite{AharonovPopescuRohrlich1,AharonovPopescuRohrlich2, CollinsPopescu}, and we present qualitatively new and subtle effects that appear in networks of quantum frames of reference, that bring into question the very nature of what conserved quantities are.

In quantum mechanics, frames of reference, whilst long ignored, are now viewed as essential in our understanding of quantum physics~\cite{AharonovPopescuRohrlich1,AharonovPopescuRohrlich2, CollinsPopescu,ChargeSSR, aharonov1967observability,AharonovKaufherr,thirdparticleparadox, krumm2021quantum,Rovelli_1991, BartlettRudolfSpekkens2, PopescuShort1, PopescuShort2, suleymanov2024nonrelativistic}, with implications across quantum foundations, quantum information, and even approaches to quantum gravity.

\begin{figure}[ht]
    \centering
    \includegraphics[width=\columnwidth]{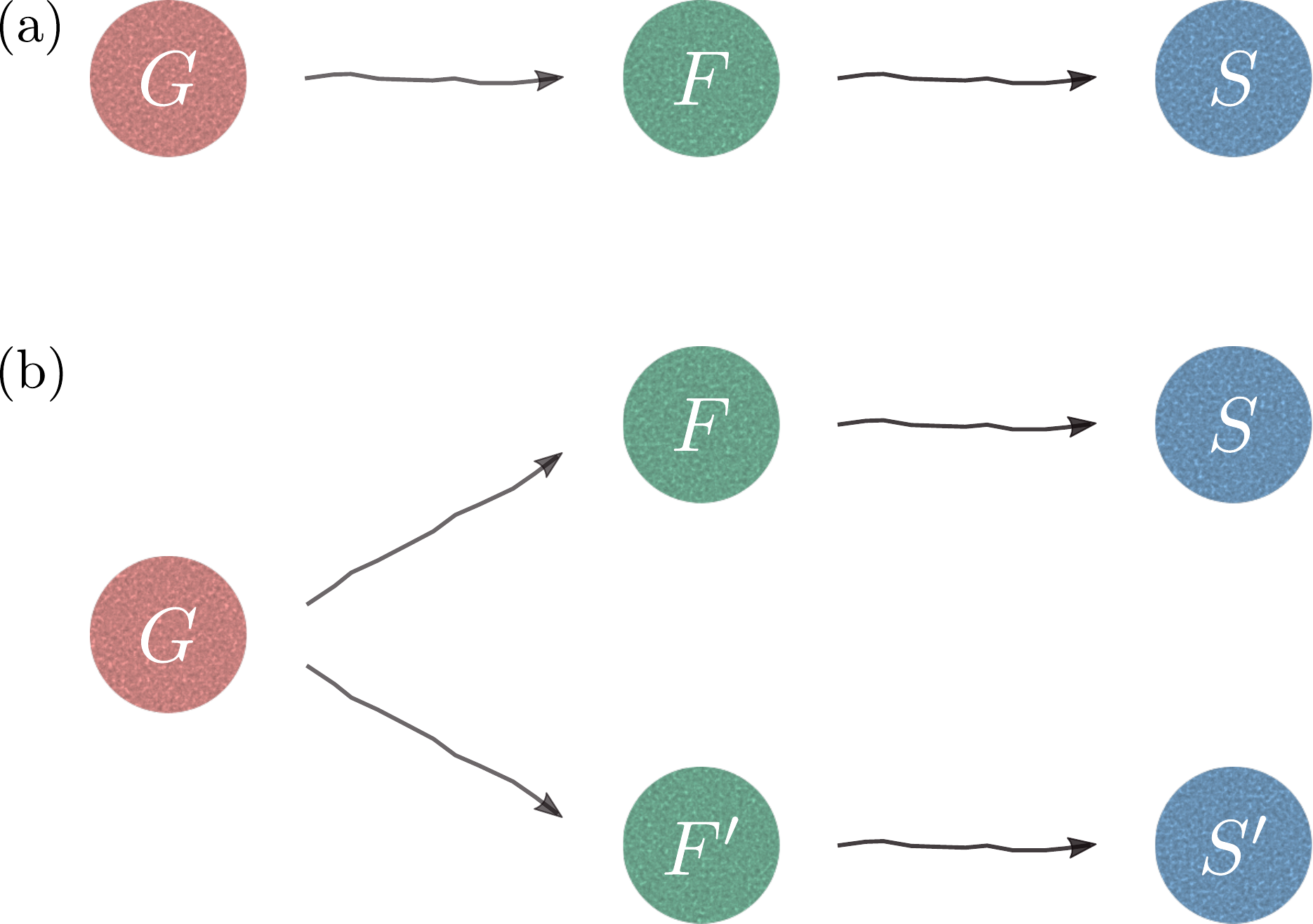}
    \caption{Chains of system preparations. (a) A grand-frame $G$ prepares a frame $F$, which then prepares a system $S$. (b) A grand-frame $G$ prepares two frames, $F$ and $F'$, which in turn prepare systems $S$ and $S'$, respectively. }
    \label{fig:GrandPreparer}
\end{figure}

However, in virtually all discussions of quantum frames of reference the set-up has been that of a single quantum frame, to which various quantum systems have been referred. On the other hand, in Nature, and very much in practice, the situation is far more complicated, with frames of reference preparing other frames of reference and so on. 
For example, a table is prepared relative to the wall and a laser is prepared relative to the table. This is a simple chain of reference systems, see Fig.~\ref{fig:GrandPreparer}(a). A more complex network is that of two tables, each prepared relative to the same wall, and having one system prepared relative to one table, and another system prepared relative to the other table, as in Fig.~\ref{fig:GrandPreparer}(b). We will call such scenarios ``networks of quantum reference frames''.  In the present paper we introduce this general idea in the context of fundamental quantum mechanics.

The second ingredient in our paper concerns conservation laws, where recently a breakthrough was made showing that conservation laws hold not only in the usual statistical sense, in terms of distributions of measurement outcomes generated over many repeated experiments, but also for individual measurement outcomes~\cite{AharonovPopescuRohrlich1, AharonovPopescuRohrlich2, CollinsPopescu}. For example, for a particle on a circle in a superposition of angular momentum eigenstates, a measurement of angular momentum gives a particular value, and whilst the usual law refers to the distribution over all possible values from many repeated experiments, Ref.~\cite{CollinsPopescu} showed that conservation still holds for a single experimental outcome, as the angular momentum of the frame of reference used to prepare the state will decrease by precisely that outcome.

While it was long known that there is an exchange of conserved quantities between a system and the frame of reference relative to which it has been prepared~\cite{AharonovKaufherr}, from the point of view of conservation laws there was never a great need to explicitly consider this. Indeed, consider, say, a system of particles evolving under a translation-invariant Hamiltonian. Then their total momentum is conserved, and there is no need to refer to any frame of reference, let alone consider exchanging momentum with it. On the contrary, for conservation in individual cases, which gives a far more detailed view on how conservation works, the frame plays an active role in the conservation, making a much deeper connection between the two.

Coming to the specific subject of our paper, the main phenomenon is the transfer of momentum between a system and the frame used to prepare it. In a previous work~\cite{CollinsPopescu} the exchange of the conserved quantity has been studied in the simple case of a chain of frames of reference, as in Fig.~\ref{fig:GrandPreparer}(a), but as we shall show here, the way conserved quantities flow in networks of quantum frames of reference such as that in Fig.~\ref{fig:GrandPreparer}(b) is qualitatively different, and raises fundamental conceptual questions about the nature of conserved quantities.

The key feature, whose unusual properties only become apparent when analysing networks of quantum frames, is related to the flow of information that accompanies the flow of conserved quantities, which exhibits a subtle quantum structure. We also introduce the notion of the first common frame of reference of two frames, which will play an important role in our results.

In addition to the main result, we also present a change of variables that we believe will be particularly suitable to discussions related to frames of reference, not only those in this paper but also in many other scenarios, and which, as far as we know, has not yet appeared in the literature.  

The remainder of this article is organised as follows.  Section~\ref{sec:paradox} describes a paradox that illustrates the qualitatively different behaviour of networks of quantum frames. Section~\ref{sec:setup} recalls the main elements of a single system and frame, as hitherto known. Section~\ref{sec:refFrameInteraction} discusses the special properties of the interaction used to prepare a system relative to a frame. Section~\ref{sec:paradoxDetail} gives the details of the paradox. Section~\ref{sec:changeVariables} introduces our alternative approach to analysing quantum reference frames in terms of ``frame of reference coordinates'', and Section~\ref{sec:singleBranch} uses it to derive the results of a linear chain from Ref.~\cite{CollinsPopescu} in a simpler fashion. Section~\ref{sec:network} solves part of the paradox. Section~\ref{sec:extendedParadox} presents a second paradox, highlighting what is unresolved from the first. Section~\ref{sec:discussion} resolves the second paradox and discusses the flow of information. Finally, Section~\ref{sec:conclusion} summarises our results.

\section{The Paradox}
\label{sec:paradox}

We will now present the basic paradox of the paper in qualitative terms. To start, consider a particle on a circle in a state of zero angular momentum, i.e. a state with no spatial features. Suppose that the particle is now prepared by a frame of reference in a state $\chi(\theta)$, with $\theta$ relative to the frame.  There must be angular momentum transfer from the frame to the particle, unless $\chi(\theta)$ is the zero angular momentum state, as the interaction is momentum conserving. However, this ``frame of reference interaction'' is special; it is not a generic total angular momentum preserving interaction. Indeed, the main result of Ref.~\cite{CollinsPopescu} was that whenever we measure the angular momentum of the particle, the angular momentum of the frame is shifted down by precisely the measurement outcome, meaning that if the original state of the frame was $\ket{\psi}_F = \sum_{l_f} c(\ell_f)\ket{l_f}_F$ and the measurement outcome of the angular momentum of the system was $l_s$, then the frame would end up in the state
\begin{equation}
    \ket{\psi-l_s}_F \coloneqq \sum_{l_f} c(\ell_f)\ket{l_f-l_s}_F.
\end{equation}
In this case, we can say that the particle got its angular momentum from its frame {\it in every individual experiment}, regardless of the outcome of the measurement. This is {\it not} a general property of angular momentum conserving interactions, but a special property of the interaction involved in preparation.

\begin{figure}
    \centering
    \includegraphics[width=\columnwidth]{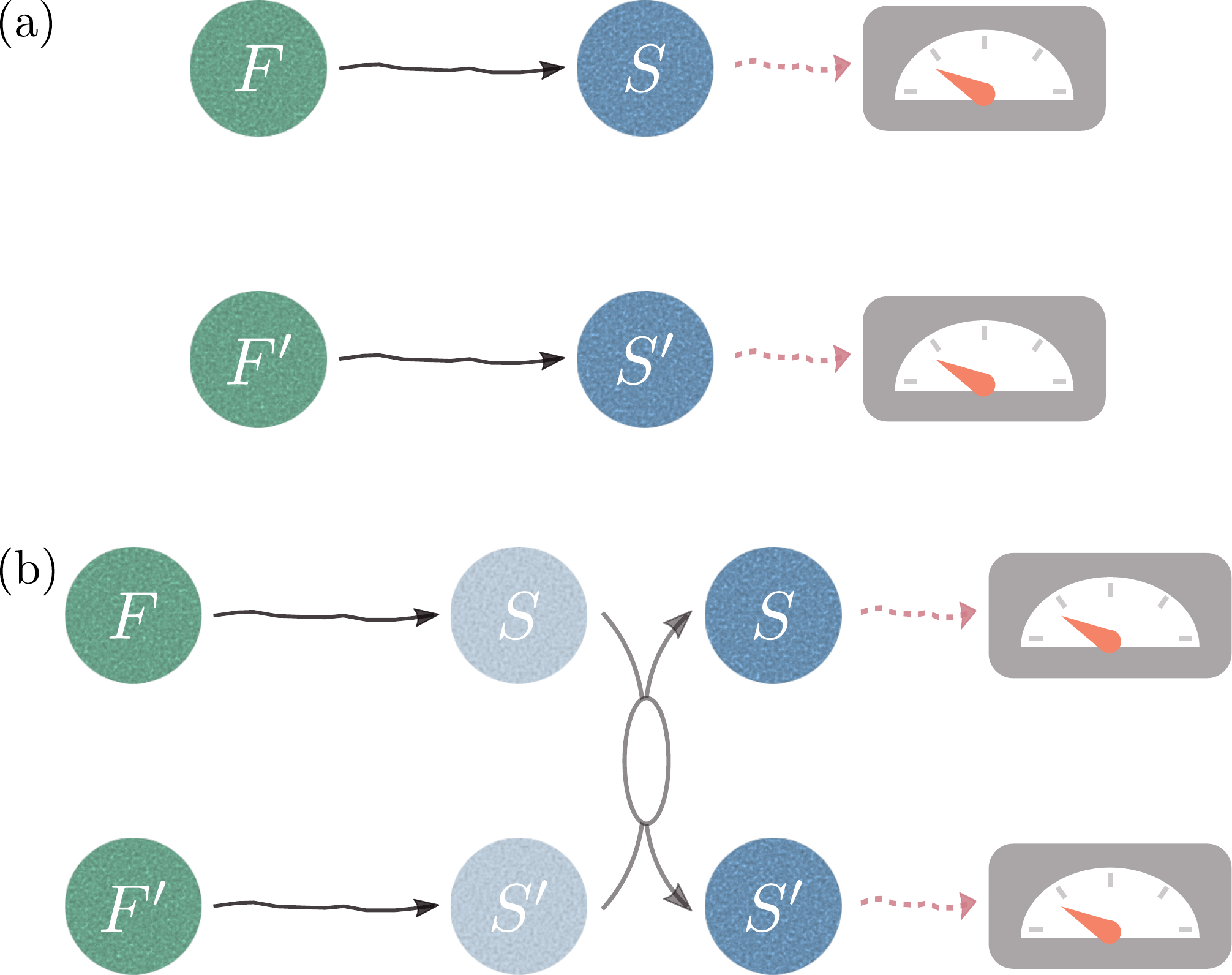}
    \caption{Simple networks of quantum frames of reference. (a) Two frames prepare each a system, which are individually measured. (b) Similar, except from the fact that the two systems interact before their measurement.}
    \label{fig:paradox}
\end{figure}

As we will show here, a paradox arises already when extending this to the simplest possible non-trivial network of frames, that of two frames, each preparing a particle as in Fig.~\ref{fig:paradox}(a).

If we measure angular momentum of each particle we will find, in every individual experiment, it came from each particle's frame. Suppose, however, that instead of measuring the angular momentum of each particle immediately after it was prepared, we first let the particles interact with each other, in an angular momentum conserving interaction, and only then we measure their angular momenta, as shown in Fig.~\ref{fig:paradox}(b). What do we expect will happen? Entering the interaction, whatever the angular momentum of a particle was, it was fully compensated by the change of angular momentum of its frame. The interaction does not change the particles' total angular momentum, only reshuffles it between them. In this case we no longer know from which frame each particle got its angular momentum, but we still expect that whatever the values of the angular momentum of the two particles, their total angular momentum must be equal to the change in the total angular momentum of the two frames.

We have found, paradoxically, that this is not the case. This is the basic paradox of this paper, and its structure and resolution will be reflected throughout our analysis of the network of frames of reference.

In particular, this raises the question of whether or not the individual conservation laws that were originally discovered in Refs.~\cite{AharonovPopescuRohrlich1,AharonovPopescuRohrlich2,CollinsPopescu}, in the context of a single frame and system, hold in general. They seem not to hold even in the simplest generalisation of two systems and two frames. As we will show, they do actually hold, provided that we take into account the way in which the two frames have been prepared to ensure their alignments are correlated with one another, such as in the branching scenario of Fig.~\ref{fig:GrandPreparer}(b). How conservation is realised in this larger set-up is by itself a very subtle mechanism, which reflects the flow of angular momentum and flow of information in frame of reference networks.

In what follows we will demonstrate the basic paradox and investigate its implications.

\section{The Setup}
\label{sec:setup}

We now review the simple chain scenario, following the basic setup discussed in Ref.~\cite{CollinsPopescu}. The main issue we wish to consider is the flow of conserved quantities between our systems of interest and their frames of reference.

Consider a particle in a superposition of various values of a conserved quantity. Using the example in Ref.~\cite{CollinsPopescu}, let it be a free particle $S$ moving on a circle in the state
\beq 
\label{initial state}
\ket{\chi}_S = \sum_{l = - \infty}^{\infty} c_{l}|\hL_S=l\ra_S,
\eeq
where $\hL_S$ is the angular momentum of the system, $l$ is an integer, and we have taken $\hbar=1$. (This example, which will also be the main set-up in the present paper, is chosen for simplicity since the eigenstates of the conserved quantity, angular momentum, are normalisable.)

Upon measuring the angular momentum, we obtain a particular outcome $\hL_S=l$, so the angular momentum of the particle changes from the initial superposition to this particular value. 
In this particular run of the experiment the angular momentum seems not to be conserved. Taking into account the measuring device (used for the ideal measurement) does not restore angular momentum conservation in this particular run, since had the initial state been an eigenstate of angular momentum, its state would not be changed by the measurement, and the measuring device would not need to exchange any angular momentum with the particle. The linearity of the Schr\"odinger equation implies that the measuring devise does not exchange angular momentum with the particle also in the case when the particle is in a superposition.  

The above is not a problem for the standard, statistical way in which conservation laws are formulated in quantum mechanics, which state that the distribution of the conserved quantity must remain unchanged. Indeed, the distribution of the observed values of angular momentum upon many runs of the experiment is the same as in the initial state, and independent on time. Yet we are left with the unpleasant fact that in an individual run angular momentum is not conserved.

The breakthrough came when it was realised~\cite{AharonovPopescuRohrlich2} that the state in Eq.~\eqref{initial state} is actually not physically well defined. Indeed, being a superposition of various values of angular momentum implies having some non-trivial angular distribution
\beq
\label{initialAngularState}
\ket{\chi}_S = \int \chi(\theta_s) \ketNoResize{\theta_s}_S d \theta_s,
\eeq
where the probability density of finding the particle at a location $\theta$, ${\cal P}(\theta)=|\chi(\theta)|^2$ is not uniform. But what is $\theta$? It cannot be an absolute position; it only makes sense relative to a frame of reference $F$.

The basic set-up is then to start with our particle of interest, $S$ in a state of angular momentum $|\hL_S=0\ra_S$ which has no preferred position, and consider a frame of reference, another particle $F$ on the circle in some state
$\psi(\theta_f)$ and use it to prepare the state of $S$ relative to it. For this reason we will alternatively refer to the frame of reference as a preparer. The initial state is
\beq
\label{initialStateFS}
\ket{\xi_0} = \ket{\psi}_F \ketNoResize{\hL_S=0}_S.
\eeq
If the intended state for $S$ is that of Eq.~\eqref{initialAngularState},
then relative to the reference frame, the state we shall actually prepare is
\beq
\label{preparedState}
\ket{\xi} = \int \psi(\theta_f) \left( \int \chi(\theta_s - \theta_f) \ketNoResize{\theta_s}_S d \theta_s \right) \ketNoResize{\theta_f}_F d \theta_f.
\eeq

If the frame of reference would be located in a precise location, the state of the particle would not be entangled with the frame, and would be in the pure state of $S$ that we originally intended. Having such a state for the frame is, however, impossible even in principle, as it would be unnormalisable. It could, however, be approximated by a strongly peaked state, leading to an infinitesimally entangled state. The surprise is that even such an infinitesimal entanglement is enough to ensure angular momentum conservation in each individual run.

In the angular momentum representation,
\beq\label{preparation in angular momentum representation}
|\xi\ra=\sum_{l_f,l_s} {\tilde\psi}(l_f){\tilde\chi}(l_s) \ket{l_f - l_s}_F \ket{l_s}_S.
\eeq
Whenever we measure the angular momentum of the particle $S$ and find it to be $l_s$, the distribution of the angular momentum of the frame $F$ conditional on the outcome $l_s$ is 
\beq
\mathcal{P}_F(l_f|\hat{L}_S=l_s) = |\tilde{\psi}(l_f + l_s)|^2.
\eeq
This is the initial distribution of the angular momentum of $F$ shifted down by $l_s$, therefore, the distribution of total angular momentum conditional on $l_s$, $\mathcal{P}(l_{total}|\hL_S=l_s)$, is equal to the initial distribution of total angular momentum, $|\tilde{\psi}(l_f)|^2$.  This ensures that conservation holds in each individual case, i.e. when conditioned on a particular outcome of the measurement on $S$. 

Another way of looking at this is that we know that the distribution of the total angular momentum of the particle and frame, $\mathcal{P}(l_{total})$, does not change since the interaction is angular momentum conserving. However, what we have proved here is that this probability conditioned on the system having value $l_s$, i.e. $\mathcal{P}(l_{total}|\hL_S=l_s)$, is independent of $l_s$.

Incidentally, the angular momentum conservation in individual cases between $F$ and $S$ is valid even when the entanglement between $F$ and $S$ is considerable and the state of $S$ is far from being pure. For this reason in the present paper we will never impose the constraint of low entanglement.

Finally, one may wonder why we claim that $\chi(\theta_s)$ is unphysical, since $\theta_s$ only makes sense relative to a frame, but take the state of the reference frame, $\psi(\theta_f)$, to be absolute. In principle, the angle $\theta_f$ only makes sense relative to a frame of its own, i.e. relative to a ``grand-frame''. It is, however, a key result of Ref.~\cite{CollinsPopescu} that, from the point of view of conservation in individual cases, it does not matter: we can treat $\theta_f$ as if it is absolute. Indeed, even if we included the grand-frame in our analysis, the conservation of angular momentum in individual cases would hold ``locally'' between the frame $F$ and the system $S$, as illustrated in Fig.~\ref{fig:net-conservation}(a). Yet, this is a highly non-trivial and fundamental issue, and how it manifests in more complex cases such as networks of quantum frames of reference is in fact a key result of this article.

\section{The frame of reference interaction}
\label{sec:refFrameInteraction}

One could be tempted to assume that once we have realised that one cannot just prepare the system in a pure state $|\chi\ra$ but one needs a frame of reference, the angular momentum conservation between them is a trivial issue. It is not. At a statistical level the conservation is indeed trivial, as the preparation of the system is an angular momentum conserving interaction. But here we have shown much more: for each outcome of a measurement of the angular momentum of the system (which started at zero), the frame is shifted by the opposite value. This is not true in general. (Just imagine a swap interaction, which is also angular momentum conserving. There the preparer ends with angular momentum zero, regardless of what the angular momentum of the system turns out to be.) The special property we found above is specific to the ``frame of reference interaction''. What is special here is that the amplitude with which the angular momentum $l_s$ is transferred between the frame $F$ and the system $S$ is independent of the initial state of the frame.

From an {\it information} point of view, what the above means is that measuring the angular momentum of the system gives us no information about what the momentum of the frame was before their interaction. All we learn is how much the angular momentum of the frame has changed.

\section{The Paradox in detail}
\label{sec:paradoxDetail}

We will now demonstrate the paradox by means of an example.  Consider two frames $F$ and $F'$ and two systems $S$ and $S'$.  Let the state of frame $F$ be
\beq
\label{initialStateOfFrame}
\ket{\psi}_F = \frac{1}{\sqrt{2}} \left(\ket{0}_F + \ket{1}_F\right),
\eeq
where $\ket{0}$ is the state with angular momentum $0$, $\ket{1}$ is the state with angular momentum $1$, and similarly for the state of $F'$. 

The initial probability distribution of the total angular momentum of the two  frames is 
\beq
\label{initProbDist}
\mathcal{P}(L_{FF'}) = \begin{cases} 1/4 : L_{FF'} = 0, \\
1/2 : L_{FF'} = 1, \\
1/4 : L_{FF'} = 2.
\end{cases}
\eeq

Let us now assume that $F$ prepares a system $S$ which is initially in a state of zero angular momentum $\ket{0}_S$, and that their joint state after the preparation is
\beq
\ket{\xi}_{FS} = \frac{1}{\sqrt{2}} \left(\ket{\psi}_F \ket{0}_S + \ket{\psi-1}_F \ket{1}_S\right),
\eeq
where $\ket{\psi-1} \coloneqq (\ket{-1} + \ket{0})/\sqrt{2}$, i.e. the state $\ket{\psi}$ decreased in angular momentum by $1$, and similar for $F'$ and $S'$. 
Note that this is a frame of reference type of interaction, with the frame $F$ acting as a frame for $S$, viewed in the momentum representation (see Eq.~\eqref{preparation in angular momentum representation}). The initial state of the four systems is then 
\beq
\begin{split}
&\ket{\xi}_{FS}\ket{\xi}_{F'S'}= \\
   & \frac{1}{2} \left(\ket{\psi,\psi}_{FF'} \ket{00}_{SS'} + \ket{\psi,\psi-1}_{FF'} \ket{01}_{SS'} \right. \\
    &\left.+ \ket{\psi-1,\psi}_{FF'} \ket{10}_{SS'} + \ket{\psi-1,\psi-1}_{FF'} \ket{11}_{SS'}\right)
\end{split}
\eeq

A measurement of the angular momentum $S$ at this stage is consistent with conservation laws in individual cases: if we measure the angular momentum of a system, its frame maintains the same probability distribution of angular momentum, only shifted to compensate for the angular momentum given to the system. 

We would like, however, to let the two systems interact via an angular momentum conserving interaction, and only after that measure their individual angular momenta.  Let the interaction between the systems be 
\beq
\label{interaction}
\begin{split}
    \hU_{SS'} &= \ket{00}\bra{00} + \frac{1}{\sqrt{2}} \left(\ket{01} + \ket{10}\right) \bra{01} \\
    &\hspace{4mm} + \frac{1}{\sqrt{2}} \left(-\ket{01} + \ket{10}\right) \bra{10} + \ket{11}\bra{11}.
\end{split}
\eeq
The state after this interaction is
\beq
\begin{split}& \hU_{SS'} \ket{\xi}_{FS}\ket{\xi}_{F'S'}=\\
    \frac{1}{2} &\Big[\ket{\psi,\psi}_{FF'} \ket{00}_{SS'} + \ket{\psi-1,\psi-1}_{FF'} \ket{11}_{SS'} \\
    &+ \frac{1}{\sqrt{2}} \left(\ket{\psi,\psi-1}_{FF'} - \ket{\psi-1,\psi}_{FF'}\right) \ket{01}_{SS'} \\
    &+ \frac{1}{\sqrt{2}} \left(\ket{\psi,\psi-1}_{FF'} + \ket{\psi-1,\psi}_{FF'}\right) \ket{10}_{SS'} \Big].
\end{split}
\eeq

Since we can no longer trace from which frame a particular system took its angular momentum, we have to look at the probability distribution of the total angular momentum of the two frames corresponding to each outcome of the measurement of the angular momenta of the two systems after the interaction.

What we expect is to see is that whatever the outcomes of the measurements of the angular momentum of each of the two systems, the distribution of the total angular momentum of the two frames is simply shifted in the opposite direction by the total of the outcomes; only in this way is the total angular momentum over the four particles preserved in each individual outcome.  In particular, suppose that after the particles interact, we find $S$ and $S'$ with angular momenta $\hL_S=0$ and $\hL_{S'}=1$ respectively. We would expect that the probability distribution of the total angular momentum of the two frames would be
\beq
\label{expectedProbDist}
\mathcal{P}(L_{FF'}) = \begin{cases} 1/4 : L_{FF'} = -1, \\
1/2 : L_{FF'} = 0, \\
1/4 : L_{FF'} = 1.
\end{cases}
\eeq

However, suppose that we find two systems, $S$ and $S'$, in the state $|01\ra_{S,S'}$.  The two frames are then in the state (after normalisation) 
\beq
\label{stateWhen01}
\frac{1}{\sqrt{6}} (\ket{\psi,\psi-1}_{FF'} - \ket{\psi-1,\psi}_{FF'}).
\eeq
The total angular momentum probability distribution of the two frames is then
\beq
\label{finalProbDist}
\mathcal{P}(L_{FF'}) = \begin{cases} 1/3 : L_{FF'} = -1, \\
1/3 : L_{FF'} = 0, \\
1/3 : L_{FF'} = 1.
\end{cases}
\eeq

Since this is different from the distribution prior to the interaction shifted down by $1$, Eq.~\eqref{expectedProbDist}, this example shows that conservation of angular momentum is no longer observed in individual cases if only the systems and frames are taken into account.

The effect is due to quantum interference.  Indeed, the two terms in Eq.~\eqref{stateWhen01} come from the fact that the post-interaction state $\ket{01}_{SS'}$ could have come from either of the pre-interaction states $\ket{01}_{SS'}$ or $\ket{10}_{SS'}$.  The two terms overlap in the angular momentum representation, in particular they both contain the term $\ket{00}_{FF'}$.  Hence the magnitude of the term $\ket{00}_{FF'}$ depends on how the two terms interfere.  Given the signs in the interaction in Eq.~\eqref{interaction}, the sign of the term in Eq.~\eqref{stateWhen01} is negative and the magnitude of the $\ket{00}_{FF'}$ is $0$. 

We have shown that in a simple example, Fig.~\ref{fig:paradox}(b), we do not have conservation in individual cases between the two frames and two systems.  This is paradoxical since if we had measured the two systems without letting them interact, we would have had conservation, and because the interaction is angular momentum preserving, apparently simply exchanging angular momentum between the two systems.  We shall spend the rest of the paper resolving this paradox.

\section{Frame of reference coordinates}
\label{sec:changeVariables}

Before continuing our discussion of the paradox, we will describe a change of variables to what we call ``frame of reference coordinates'' (FRC).  These will be useful in all discussions about frames of reference, in a similar way to how centre of mass/relative coordinates have been very useful in other situations.  While simple and natural, it also has some counterintuitive elements, and we have not encountered it before.

Since we prepare our system $S$ relative to the reference frame $F$, it is natural to change to the reference frame coordinates
\beq
\label{variables change}
\begin{split}
\htheta_1 &= \htheta_F,\\
\htheta_2 &= \htheta_S - \htheta_F,
\end{split}
\eeq
where we have taken all angles to be defined as $0 \le \htheta < 2 \pi$, and the equations with angles are taken implicitly to be $\bmod \, 2 \pi$, as we do throughout this paper.  This leads automatically to the conjugate variables
\beq
\begin{split}
\hL_1 &= \hL_F + \hL_S, \\
\hL_2 &= \hL_S,
\end{split}
\eeq
which satisfy the usual angular momentum commutation relations: 
\beq
\begin{split}
[\hL_j,e^{i\htheta_k}] &= \delta_{jk} e^{i\htheta_k}, \\
[\htheta_j,\htheta_k] &= 0, \\
[\hL_j,\hL_k] &= 0,
\end{split}
\eeq
for all pairs $(j,k)$.  Therefore $\hL_1$ and $\hL_2$ are the ``canonical conjugate'' angular momenta on the circle associated with the new angles $\htheta_1$ and $\htheta_2$.

A caveat: Since we are talking about angles and angular momentum, when changing variables one should be more careful than when discussing linear position and momentum, since angles should be defined between $0$ and $2\pi$ and angular momentum can only have integer quantities.  For an example of what could go wrong, see Appendix~\ref{AppendixBadCoordinates}.

As one could readily check, the original Hilbert space of $S$ and $F$,  $\mathcal{H}=\mathcal{H}_F \otimes \mathcal{H}_S$, can now be viewed as the  tensor product $\mathcal{H}=\mathcal{H}_1 \otimes \mathcal{H}_2$. Therefore we can talk about the entanglement between ``particles'' $1$ and $2$, local operations on each ``particle'', and measurements on each ``particle'' which leave the other ``particle'' unchanged, just as we can with the usual coordinates for any real particles.

In Dirac notation this leads to a change from the original Hilbert space decomposition $\mathcal{H}=\mathcal{H}_F \otimes \mathcal{H}_S$, where the state is written as
\beq
\label{stateInSFCoords}
\int \psi(\theta_f) \chi(\theta_s - \theta_f) \ket{\theta_s,\theta_f} d \theta_s d \theta_f,
\eeq
to the new Hilbert space decomposition $\mathcal{H}=\mathcal{H}_1 \otimes \mathcal{H}_2$, where the state is written
\beq
\int \psi(\theta_1) \chi(\theta_2) \ket{\theta_1,\theta_2} d\theta_1 d\theta_2 = \ket{\psi}_1 \ket{\chi}_2.
\eeq
To prove that this is how the state transforms under a change of coordinates, we act on Eq.~\eqref{stateInSFCoords} with the identity
\beq
\id = \int \ket{\theta_1,\theta_2} \bra{\theta_1,\theta_2} d \theta_1 d \theta_2,
\eeq
and use the inner product
\beq
\braket{\theta_1,\theta_2}{\theta_f,\theta_s} = \delta(\theta_1 - \theta_f) \delta(\theta_2 - \theta_s + \theta_f).
\eeq

What makes this change of variables particularly effective in frame of reference problems is how it behaves under Fourier transforms, i.e. when going from angles to their conjugate momenta:
\beq
\label{coordTransforms}
\centering
\begin{array}{ccc}
\psi(\theta_f) \chi(\theta_s - \theta_f) & \xleftrightarrow{coord} & \psi(\theta_1) \chi(\theta_2)\\
\\
\mathcal{F} \downarrow  \uparrow \mathcal{F}^{-1}  &  & \mathcal{F} \downarrow  \uparrow \mathcal{F}^{-1}\\
\\
\tilde{\psi}(l_f + l_s) \tilde{\chi}(l_s) & \xleftrightarrow{coord} & \tilde{\psi}(l_1) \tilde{\chi}(l_2),\\
\end{array}
\eeq
where $\tilde{\psi}(l)$ and $\tilde{\chi}(l)$ are the Fourier transforms of $\psi(\theta)$ and $\chi(\theta)$ (see Appendix~\ref{AppendixFourierTransform} for a proof of the Fourier transform from $(\htheta_F,\htheta_S)$ to $(\hL_F,\hL_S)$). Note the simplicity of the transformation in the new variables, where the state is a product state and the ``particles'' $1$ and $2$ transform from angles to angular momentum separately.

Furthermore, and most important, we would like to emphasise that $\psi$ and $\chi$ are the same functions in both the old and the new variables. This makes the conversion extremely powerful.

Finally, a note on what is counterintuitive. Take for example the variables $\htheta_F$ and  $\htheta_1$. According to their definition in Eq.~\eqref{variables change}, they actually denote the same variable---the position of the frame. However, when considering their canonical conjugate (and the associated Fourier transform), it depends on which Hilbert space tensor partition we consider: it is $\hL_F$ in the $\mathcal{H}=\mathcal{H}_F \otimes \mathcal{H}_S$ decomposition, but $\hL_1=\hL_F+\hL_S$ in the $\mathcal{H}=\mathcal{H}_1 \otimes \mathcal{H}_2$ decomposition.

\section{A simple chain of frames of reference}
\label{sec:singleBranch}

Before addressing the basic paradox, it is useful to gain further insights into the flow of conserved quantities in a simpler situation, that of the linear chain of Fig.~\ref{fig:GrandPreparer}(a), using our new tools.

The linear chain has been considered in Ref.~\cite{CollinsPopescu} in order to answer a basic question. Given that even a small amount of entanglement between a frame of reference and a system is enough to ensure conservation in individual cases, the question arose of whether this leads us into an infinite regression, having to consider frames of frames and so on until eventually having to consider the whole universe. This would have made the whole idea of conservation physically irrelevant. A central result of Ref.~\cite{CollinsPopescu}, working in a linear chain of frames of reference as in Fig.~\ref{fig:GrandPreparer}(a), was that this is not the case: we only need to consider the first link of the chain. That is, we do not have to consider the frame of the frame, or anything else.

We shall start with the basic setup containing just a frame and a system, as discussed in the introduction, and show that angular momentum is conserved for individual outcomes when we measure $S$. Afterwards we shall add another frame at the beginning of the chain, $G$, and show that it does not participate in the conservation for measurements on $S$.

Using the change of variables from the previous section, the results of Ref.~\cite{CollinsPopescu} follow easily, as we shall now demonstrate.

Instead of having Eqs.~\eqref{initialStateFS} and~\eqref{preparedState}, we shall have a much simpler equation representing the preparation of the system,
\beq
\label{preparationIn12}
\ket{\xi_0} = \ket{\psi}_1 \ketNoResize{\hL_2 = 0}_2 \xrightarrow{\text{prepare}} \ket{\xi} = \ket{\psi}_1 \ket{\chi}_2.
\eeq

The advantage of this is that whilst the prepared state is entangled in the $F$, $S$ basis, it is a product state in the $1$, $2$ basis. In these coordinates, the preparation acts only in $\mathcal{H}_2$, leaving $\mathcal{H}_1$, which consists of the angle of the preparation device and the total momentum, unchanged, and this allows us to understand the results in a much easier way.

To prove the above, we begin by transforming the initial state of Eq.~\eqref{initialStateFS}, $\ket{\psi}_F \ketNoResize{\hL_S=0}_S$, into the $1$, $2$ coordinates. We note that in the basis $(\theta_s,\theta_f)$, this initial wavefunction has the form $\psi(\theta_f) \chi_0(\theta_s - \theta_f)$ with $\chi_0 = 1/\sqrt{2 \pi}$.  Hence we can use the top line of Eq.~\eqref{coordTransforms} to transform it into the angular basis $(\theta_1, \theta_2)$, which gives $\psi(\theta_1) \chi_0(\theta_2)$, i.e. the state
\beq
\ket{\xi_0} = \ket{\psi}_1 \ketNoResize{\hL_2 = 0}_2.
\eeq

We can write the state after the system is prepared, Eq.~\eqref{preparedState}, in the $1$,$2$ coordinates by using Eq.~\eqref{coordTransforms}, giving
\beq
\ket{\xi} = \iint \psi(\theta_1) \chi(\theta_2) \ket{\theta_1}_1 \ket{\theta_2}_2 d \theta_1 d \theta_2
= \ket{\psi}_1 \ket{\chi}_2,
\eeq
which completes the proof of Eq.~\eqref{preparationIn12}.  

After preparing our system, we measure $\hL_S$, the angular momentum of $S$, whilst leaving $F$ unchanged: suppose the outcome is $l_s$. Then, since $\hL_2 = \hL_S$, the value of $\hL_2$ is also $l_s$, and the state of the system collapses onto
\beq
\ket{\psi}_1 \ket{l_s}_2.
\eeq
We can write the entire series of events compactly as
\beq
\begin{alignedat}{2}
\ket{\xi_0} & \overset{\hphantom{measure}}{=} &&\ket{\psi}_F \ket{0}_S \\
& \overset{\hphantom{measure}}{=} &&\ket{\psi}_1 \ket{0}_2 \\
& \xrightarrow{\text{prepare}} &&\ket{\psi}_1 \ket{\chi}_2 \\
& \xrightarrow{\text{measure}} &&\ket{\psi}_1 \ket{l_s}_2,
\end{alignedat}
\eeq
where $\ket{0}$ is shorthand for zero angular momentum, i.e. the state $\ketNoResize{\hL = 0}$.
To interpret the meaning of this final state, we note that we can rewrite it as
\beq
\ket{\psi}_1 \ket{l_s}_2 = e^{i \htheta_2 l_s} \ket{\psi}_1 \ket{0}_2
= e^{i \htheta_S l_s} e^{- i \htheta_F l_s} \ket{\xi_0},
\eeq
where $e^{i \htheta l}$ is the shift operator which increases the angular momentum by $l$, and where in the last equality we went back to the original variables $S$, $F$ which are of ultimate interest for us.  The final state is equal to the original state with the angular momentum of $S$ increased by $l_s$ and the angular momentum of $F$ decreased by $l_s$.  Hence total angular momentum is conserved for each individual measurement outcome.  

To answer the question of who prepared the frame, and whether that needs to be taken into account, in Ref.~\cite{CollinsPopescu} a grand-frame $G$ was introduced, and it was shown that the grand-frame does not participate in the conservation of the angular momentum of the system.  This scenario is represented in Fig.~\ref{fig:net-conservation}(a). We now prove that result in a simplified way.

We start with the grand-frame in the state $\ket{\phi}_G$, and the frame and the system both in the state with no angular dependence, i.e. with angular momentum zero:
\beq
\ket{\Phi_0} = \ket{\phi}_G \ket{0}_F \ket{0}_S.
\eeq
First we prepare $F$ relative to $G$, similar to how we prepared $S$ relative to $F$ in Eq.~\eqref{preparedState}, which gives
\beq
\ket{\Phi} = \int \phi(\theta_g) \psi(\theta_f - \theta_g) \ketNoResize{\theta_g}_G \ketNoResize{\theta_f}_F \ket{0}_S d \theta_g d \theta_f .
\eeq
Next we prepare $S$ relative to $F$, which gives:
\beq
\int \phi(\theta_g) \psi(\theta_f - \theta_g) \chi(\theta_s - \theta_f)   \ketNoResize{\theta_g}_G \ketNoResize{\theta_f}_F \ketNoResize{\theta_s}_S  d \theta_g d \theta_f d \theta_s.
\eeq

We shall change to coordinates where $F$ is relative to $G$ and $S$ is relative to $F$:
\beq
\begin{alignedat}{3}
\htheta_0 &= \htheta_G \; &&; \; \hL_0 &&= \hL_G + \hL_F + \hL_S \\
\htheta_1 &= \htheta_F - \htheta_G \; &&; \; \hL_1 &&= \hL_F + \hL_S \\
\htheta_2 &= \htheta_S - \htheta_F\; &&; \; \hL_2 &&= \hL_S.
\end{alignedat}
\eeq
These have the usual commutation relations and map one-to-one with the original coordinates, and the Hilbert space can be split into $\mathcal{H}_0 \otimes \mathcal{H}_1 \otimes \mathcal{H}_2$.  $0$, $1$ and $2$ can be thought of as separate particles as before. 

Using these coordinates, the initial state is
\beq
\ket{\Phi_0} = \ket{\phi}_0 \ket{0}_1 \ket{0}_2,
\eeq
and our experiment is described by
\beq
\label{grandPreparer}
\begin{alignedat}{2}
\ket{\phi}_G \ket{0}_F \ket{0}_S 
& \overset{\hphantom{measure \,\hL_S}}{=} &&\ket{\phi}_0 \ket{0}_1 \ket{0}_2 \\
& \xrightarrow{\text{prepare F}} && \ket{\Phi} = \ket{\phi}_0 \ket{\psi}_1 \ket{0}_2\\
& \xrightarrow{\text{prepare S}} &&\ket{\phi}_0 \ket{\psi}_1 \ket{\chi}_2\\
& \xrightarrow{\text{measure }\hL_S} &&\ket{\phi}_0 \ket{\psi}_1 \ket{l_s}_2\\
& \overset{\hphantom{measure \,\hL_S}}{=} && e^{i \htheta_2 l_s} \ket{\phi}_0 \ket{\psi}_1 \ket{0}_2\\
& \overset{\hphantom{measure \,\hL_S}}{=} && e^{i \htheta_S l_s} e^{- i \htheta_F l_s} \ket{\Phi},
\end{alignedat}
\eeq
where $\ket{\Phi}$ is the state of all particles immediately after $F$ has been prepared.

The final state, corresponding to measuring the system and finding the value $l_s$, is the state immediately after the preparation of $F$, $\ket{\Phi}$, with the angular momentum of $S$ increased by $l_s$, and the angular momentum of $F$ decreased by $l_s$.  In the new coordinates, which give a much simpler view, the entire complexity of the state in its original variables is not apparent.  For example, $G$ and $F$ are entangled immediately after the preparation of $F$.  For our purpose however, this doesn't matter as the flow of angular momentum can be readily deduced.  Since the shifts of angular momentum $e^{i \htheta_S l_s} e^{- i \htheta_F l_s}$ are unitary operations acting on $F$ and $S$, they do not change $G$. Therefore the angular momentum of the system comes directly from $F$: there is no need to include $G$ when discussing conservation of $S$ in individual cases.

\section{A network of reference frames}
\label{sec:network}

The basic paradox involved two frames of reference, $F$ and $F'$, each of which prepared their own system. We will now revisit the idea that the angles of these frames, $\theta_f$ and $\theta_{f'}$, are not absolute, but that they only have meaning relative to some grand-frame.  For a single particle and a single frame, we have shown that there is no need to consider the grand-frame of the frame, but here we will show that the situation is different. Indeed, the basic paradox has been formulated in the angular momentum representation, however, since it ultimately depends upon interference between various angular momentum terms in the states of the frames, their angles also play a role. 

\begin{figure*}
    \centering
    \includegraphics[width=.8\textwidth]{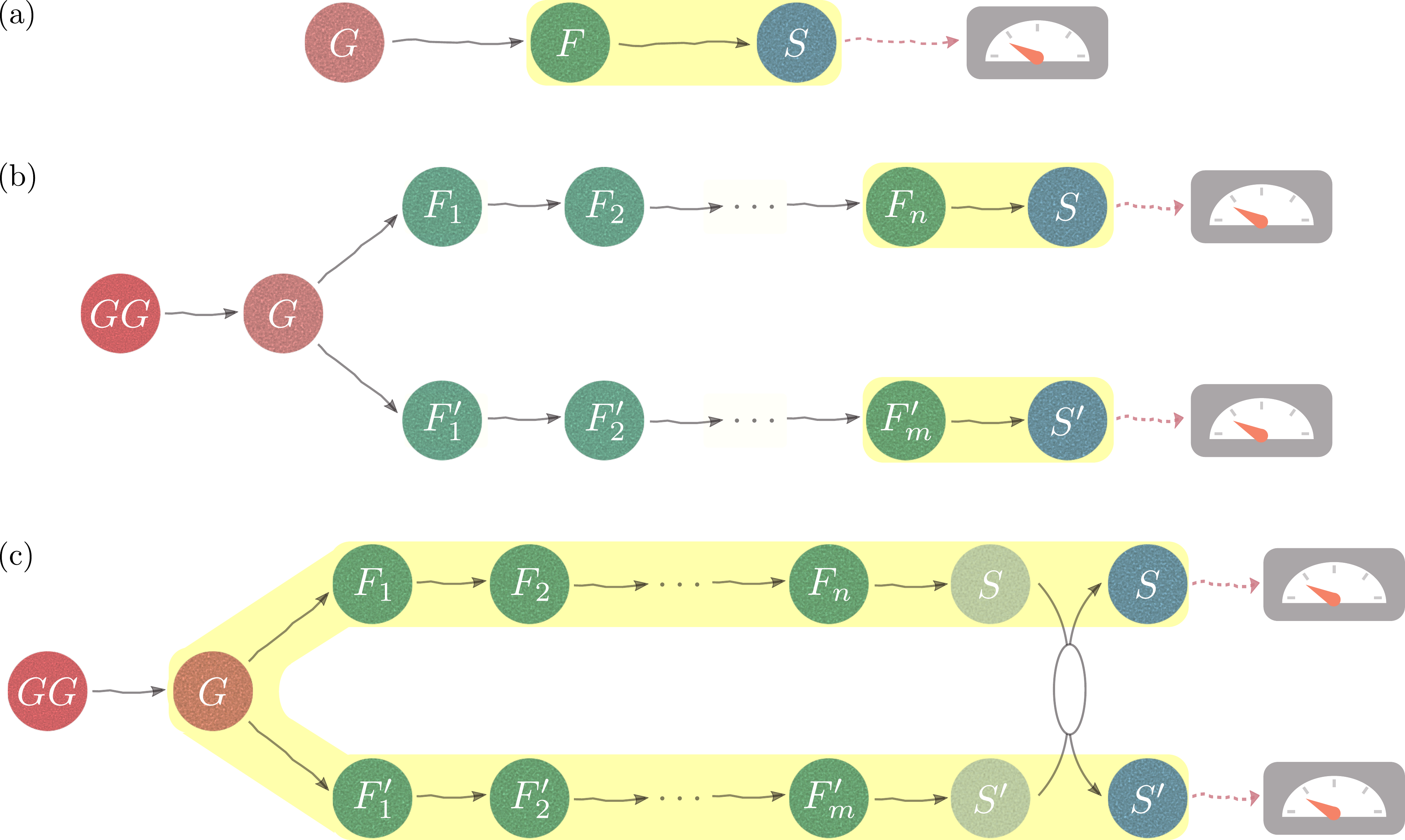}
    \caption{Conservation laws in networks of quantum frames of references.  Note that the proofs in the text consider only a single $F$ and a single $F'$, but they apply in full generality. (a) In a simple chain, the conservation law for individual outcomes is local between the measured system and its frame of reference.  (b) In a network, so long as there is no interaction between different branches, the conservation law for individual outcomes remains localised. (c) However, if systems in distinct branches interact before their measurement, the conservation law for individual outcomes requires the inclusion of every frame up to and including their first common grand-frame in the network.}
    \label{fig:net-conservation}
\end{figure*}

It is natural then to consider a network of frames of reference, such as the branching frame illustrated in Fig.~\ref{fig:net-conservation}(b) and (c), where we have a common grand-frame $G$ which is used to prepare the frames $F$ and $F'$.  Here $G$ is the first common frame for all our systems of interest.  As we will show in the following sections, once we have this full structure, the individual conservation laws, which would seem to be violated when considering two frames, are restored, albeit by an equally puzzling effect.  

What we will show is that:
\begin{enumerate}[label=\Alph*.]
\item If we measure the angular momenta of the systems $S$ and $S'$ before they interact with one another, we will find, similar to the previous section, that conservation holds in each individual case between the systems and the frames $F$ and $F'$, and the state of $G$ does not change and hence $G$ plays no role in the conservation. 
\item If instead we let the systems interact with one another using an angular momentum conserving unitary operation, and then measure their angular momenta, we will show that angular momentum is no longer conserved between $S$, $S'$, $F$ and $F'$ in individual cases.  
\item In the case where the systems interact with one another, we will show that angular momentum is conserved in individual cases if we also include $G$ in the total.  The state of $G$ changes, and it is now essential to include it in order to have this conservation.
\item An infinite regression of frames is not needed: any other frames prior to $G$, which is the first common frame of reference for all the other systems we are interested in,  do not participate in the transfer of angular momentum needed for conservation in individual cases.
\end{enumerate}

\subsection{Branching chain of frames with no interactions in between}
\label{sec:branchingChain}

We begin with $G$ in the state $\ket{\phi}_G$, and the other frames and systems in the state with angular momentum zero:
\beq
\ket{\Psi_0} = \ket{\phi}_G \ket{0}_F \ket{0}_{F'} \ket{0}_S \ket{0}_{S'}.
\eeq
We prepare $F$ and $F'$ relative to $G$, similar to Eq.~\eqref{preparedState}, which gives the state 
\beq
\label{twoPreparersState}
\begin{split}
\ket{\Psi} = &\int \phi(\theta_g) \psi(\theta_f - \theta_g) \psi'(\theta_{f'} - \theta_g) \ket{0}_S \ket{0}_{S'} \\
& \quad \; \ket{\theta_g}_G \ket{\theta_f}_F \ket{\theta_{f'}}_{F'} d \theta_g d \theta_f d \theta_{f'}.
\end{split}
\eeq 

Then we prepare $S$ relative to $F$ and $S'$ relative to $F'$, which gives the state 
\beq
\label{twoSystemsPreparedState}
\begin{split}
\ket{\Phi} = &\int \phi(\theta_g) \psi(\theta_f - \theta_g) \psi'(\theta_{f'} - \theta_g) \chi(\theta_s - \theta_f) \chi'(\theta_{s'} -\theta_{f'}) \\
& \quad \; \ket{\theta_g}_G \ket{\theta_f}_F \ket{\theta_{f'}}_{F'} \ket{\theta_s}_S \ket{\theta_{s'}}_{S'} d \theta_g d \theta_f d \theta_{f'} d \theta_s d \theta_{s'}.
\end{split}
\eeq

In order to simplify our analysis, we shall use the coordinates
\beq
\label{twoPreparerCoordinates}
\begin{alignedat}{6}
&\htheta_A &&= \htheta_G && &&; \, &&\hL_A &&= \hL_G  +  \hL_F  +  \hL_{F'}  +  \hL_{S}  +  \hL_{S'}\\
&\htheta_B &&= \htheta_F &&- \htheta_G &&; \; \; && \hL_B &&= \hL_F  +  \hL_{F'}  +  \hL_{S}  +  \hL_{S'}\\
&\htheta_R &&= \htheta_{F'}  &&- \htheta_F &&; && \hL_R &&= \hL_{F'}  +  \hL_{S'}\\
&\htheta_C &&= \htheta_{S} &&- \htheta_F &&; && \hL_C &&= \hL_{S} \\
&\htheta_{C'} &&= \htheta_{S'} &&- \htheta_{F'} &&; && \hL_{C'} &&= \hL_{S'},
\end{alignedat}
\eeq
where the angles are defined $\bmod \: 2 \pi$ as usual.  We can think of $A,B,R,C$ and $C'$ as independent particles, and the total Hilbert space is a tensor product of Hilbert spaces of these ``particles''. $\hL_A$ is the total momentum of all particles, and $\hL_B$ is the total angular momentum excluding $G$. 

In these coordinates the state immediately after $F$ and $F'$ are prepared, Eq.~\eqref{twoPreparersState}, is written, by substituting the angles from Eq.~\eqref{twoPreparerCoordinates}, as
\beq
\ket{\Psi} = \ket{\phi}_A \ket{\eta}_{BR}  \ket{0}_C  \ket{0}_{C'},
\eeq
where
\beq
\ket{\eta}_{BR} = \int  \psi(\theta_b) \psi'(\theta_r + \theta_b)  \ket{\theta_b}_B  \ket{\theta_r}_R  d \theta_b d \theta_r.
\eeq

This state is entangled between $B$ and $R$, but these are in a product state with $A$, $C$ and $C'$.

In our new coordinates the state after preparation of the two systems, Eq.~\eqref{twoSystemsPreparedState}, becomes
\beq
\label{statePriorToInteractionAngle}
\ket{\Phi} = \ket{\phi}_A \ket{\eta}_{BR} \ket{\chi}_C \ket{\chi'}_{C'}.
\eeq

If we were to measure the two systems at this point, and found $\hL_S = l_s$ and $\hL_{S'} = l_{s'}$, the state would become:
\beq
\begin{split}
& \ket{\phi}_A \ket{\eta}_{BR} \ket{l_s}_C \ket{l_{s'}}_{C'} \\ 
& = e^{i \htheta_C l_s} e^{i \htheta_{C'} l_{s'}} \ketNoResize{\Psi} \\
& = e^{i \htheta_S l_s} e^{-i \htheta_F l_s} e^{i \htheta_{S'} l_{s'}} e^{-i \htheta_{F'} l_{s'}} \ketNoResize{\Psi},
\end{split}
\eeq
where in the last line we transformed back to the coordinates we are most interested in, those of $F$, $S$, etc., using Eq.~\eqref{twoPreparerCoordinates}. This is the wavefunction prior to preparing the systems, $\ketNoResize{\Psi}$, with $l_c$ moved from $F$ to $S$, and $l_{c'}$ moved from $F'$ to $S'$. Hence conservation holds in these individual cases involving only $S$, $F$, $S'$ and $F'$, without $G$. Conservation in each branch occurs exactly the same as it did in the simple $G-P-S$ chain of Fig.~\ref{fig:GrandPreparer}.

\subsection{Branching chain of frames with system-system interaction}
\label{sec:paradoxWithG}

In the basic paradox described in Section~\ref{sec:paradox}, we showed that conservation did not hold in individual cases in a specific example with two frames and two systems, where the angular momentum of the two systems was measured after they interact using an angular momentum conserving interaction.  One may wonder whether this non-conservation is an artifact due to not explicitly taking into account the way in which the two frames were prepared to be aligned with one another.  In this section, we shall show that even if we include in our model a grand-frame which is used to align the two frames, as in Fig.~\ref{fig:GrandPreparer}(b), we still do not have conservation in individual cases between the two frames and two systems alone. 

We do this by an example which is an extension of the basic paradox example in Section~\ref{sec:paradoxDetail}, the difference being that here we have an explicit grand-frame $G$.  We shall begin with all systems in the state where everything is in the state with angular momentum zero:
\beq
\ket{0}_G \ket{0}_F \ket{0}_{F'} \ket{0}_S \ket{0}_{S'}.
\eeq

We shall then use a frame of reference interaction to prepare $F$ relative to $G$, giving
\beq
\frac{1}{\sqrt{2}} \left(\ket{0}_G \ket{0}_F + \ket{-1}_G \ket{1}_F \right) \ket{0}_{F'} \ket{0}_S \ket{0}_{S'},
\eeq
which is the equivalent of the initial state of $F$ in Eq.~\eqref{initialStateOfFrame} in Section~\ref{sec:paradoxDetail}, i.e. of  $\frac{1}{\sqrt{2}}(\ket{0}_F + \ket{1}_F)$, but this time relative to $G$.  We then prepare $F'$ relative to $G$ in a similar way, which gives:
\beq
\frac{1}{2} \sum_{m,n = 0}^1 \ket{-m-n}_G \ket{m}_F \ket{n}_{F'} \ket{0}_S \ket{0}_{S'}.
\eeq
The distribution of total angular momentum of the two frames at this point is 
\beq
\label{initProdDistWithG}
\mathcal{P}(L_{FF'}) = \begin{cases} 1/4 : L_{FF'} = 0, \\
1/2 : L_{FF'} = 1, \\
1/4 : L_{FF'} = 2.
\end{cases}
\eeq

Next we use similar interactions to prepare $S$ relative to $F$ and $S'$ relative to $F'$, giving the state
\beq
\frac{1}{4} \sum_{m,n,k,l = 0}^1 \ket{-m-n}_G \ket{m-k}_F \ket{n-l}_{F'} \ket{k}_S \ket{l}_{S'}.
\eeq

We know, as proven in Section~\ref{sec:branchingChain}, that at this moment the angular momentum of each system and its frame is conserved in each individual case.

Then we interact the systems $S$ and $S'$, using the interaction of Eq.~\eqref{interaction}, and finally measure $S$ and $S'$.  If we find the outcome $\ket{0}_S \ket{1}_{S'}$, the state of the grand-frame and frames is
\beq
\frac{1}{\sqrt{6}} \sum_{m,n = 0}^1 \ket{-m-n}_G ( \ket{m}_F \ket{n-1}_{F'} 
- \ket{m-1}_F \ket{n}_{F'} ).
\eeq

As in our previous example, we have destructive interference in the term with $\ket{0}_F \ket{0}_{F'}$, and the distribution of total angular momentum of the two frames is now 
\beq
\mathcal{P}(L_{FF'}) = \begin{cases} 1/3 : L_{FF'} = -1, \\
1/3 : L_{FF'} = 0, \\
1/3 : L_{FF'} = 1.
\end{cases}
\eeq
This is not the initial distribution of Eq.~\eqref{initProdDistWithG} shifted down by the total angular momentum of $S$ and $S'$, i.e. shifted down by $1$, hence the probability distribution of the total angular momentum of the two frames and the two systems has changed, and, just as in Section~\ref{sec:paradoxDetail}, we no longer have conservation in this individual case between the two frames and two systems alone. 

On the other hand, it follows directly that once the grand-frame is included in the total, angular momentum is conserved in every individual case. Indeed, since the total angular momentum of all parties is an eigenstate of zero angular momentum, we will find that in every individual case the total angular momentum is zero. Therefore, in this example, there is conservation in individual cases, so long as we include the grand-frame in the total. Indeed, the grand-frame is essential in order to have individual case conservation.

\subsection{The grand-frame provides conservation}

In the previous section we showed that we restored the conservation in a specific example by including the grand-frame. We now prove this in full generality.  

First we note that a general interaction between $S$ and $S'$ which preserves total angular momentum may depend on $\htheta_S - \htheta_{S'}$, $\hL_S$, and $\hL_{S'}$, but not on $\htheta_S + \htheta_{S'}$, as that would change the total angular momentum. Therefore our interaction may be written as $\hU(\htheta_S - \htheta_{S'}, \hL_S, \hL_{S'})$. 

In our new coordinates, defined in Eq.~\eqref{twoPreparerCoordinates}, the relative angle $\htheta_S - \htheta_{S'}$ becomes
\beq
\htheta_{rel} = \htheta_C - \htheta_{C'} + \htheta_R,
\eeq
which is taken $\bmod \, 2 \pi$ as usual. We write the interaction $\hU(\htheta_{rel}, \hL_C, \hL_{C'})$ as $\hU_{RCC'}$, where we have added the indices $RCC'$ to denote the systems on which it acts. The state is now given by applying $\hU_{RCC'}$ on the state after the preparation of the systems, $\ket{\Phi}$ from Eq.~\eqref{statePriorToInteractionAngle}, which gives 
\beq
\hU \ket{\Phi} = \ket{\phi}_A  \hU_{RCC'} \ket{\eta}_{BR} \ket{\chi}_C \ket{\chi'}_{C'}.
\eeq

Finally we measure $\hL_S$ and $\hL_{S'}$, the angular momenta of $S$ and $S'$, and find some particular values $l_s$ and $l_{s'}$. Since $\hL_C = \hL_S$ and $\hL_{C'} = \hL_{S'}$, Eq.~\eqref{twoPreparerCoordinates}, this means $\hL_C = l_s$ and $\hL_{C'} = l_{s'}$, and the state becomes (up to normalisation)
\beq
\ket{\phi}_A \ket{l_s}_C \ket{l_{s'}}_{C'} \braL{C}{l_s} \braL{C'}{l_{s'}} \hU_{RCC'} \ket{\eta}_{BR} \ket{\chi}_C \ket{\chi'}_{C'}.
\eeq

Since system $A$, whose angular momentum $\hL_A$ is equal to the total momentum of all particles (including $G$), remains in a direct product state with everything else (i.e. with the state describing the particular distribution of the angular momenta between the different particles), the total momentum remains unchanged whatever the measured values of the angular momenta of the systems, $l_s$ and $l_{s'}$. Hence we do have momentum conservation in individual cases if we consider all five particles $GFF'SS'$ together.

\subsection{Is infinite regression now needed?}

We have just shown that there are scenarios, such as Fig.~\ref{fig:net-conservation}(c), where we need to go further back in the chain than the immediate frames in order to have conservation in individual cases.  This is completely different to the linear case of Fig.~\ref{fig:GrandPreparer}(a), where the immediate frame was enough.   Does this mean that we need to take into account the infinite regression of great-grand frames and so on?  

It is easy to prove that in the branching case everything stops after the grand-frame.  In other words, we need to include all systems back to and including the first common frame of reference, and there is no need to go back any further.  The conservation law in individual cases is confined, and does not have infinite regression.

\section{The extended paradox}
\label{sec:extendedParadox}

By adding the first common frame of reference, $G$, to the basic paradox of Section~\ref{sec:paradox}, and making all the angles relative, the individual angular momentum conservation law has been restored. One might be tempted to think that the basic paradox has been a simple artifact of not defining the angles properly and that the paradox is by now solved. However, a paradox still remains: how does the grand-frame, which was not involved in conservation in individual cases prior to the interaction between particles, become involved in the conservation after such an interaction, as in Fig.~\ref{fig:net-conservation}(c)?

To better understand what the problem is, let us delve deeper into understanding the flow of information about the conserved quantities, using the classical analogue of this scenario.

The situation seems similar to the following: Alice asks her local bank branch for a loan of gold coins. However, instead of asking for a precise amount, she asks the bank to give her an amount equal to the result of a die throw. The bank agrees to honour the request, regardless of how many coins it has in its vault (an amount which is far larger than the maximal Alice's request). The bank branch originally got the coins in its vault from the central bank, according to a similar protocol. Knowing how much money Alice got only tells us by how much the bank branch reserve decreased, but it is totally uncorrelated to how much the central bank's reserve decreased when it supplied the branch. Hence, the coin conservation is restricted to the Alice and her bank branch. This is {\it not} a trivial consequence of the total coin conservation. Indeed, if Alice would ask her branch to lend her $1\%$ of the money in its vault, the amount she receives will be correlated with the amount the central bank transferred to the branch. 

In a similar way, we could now consider Alice and Bob, each requesting a lone from their (separate) branches of the same central bank. Knowing how much each of them received will still be uncorrelated with how much the central bank gave the two branches. 

Finally, suppose that Alice and Bob decide to exchange some coins, again by a probabilistic protocol. It is clear that knowing at the end of this process how many coins Alice and Bob have, does not give any further information about by how much the central  bank reserve decreased. Yet, in our case, the angular momentum - the equivalent of coins - becomes correlated with the decrease in the angular momentum of the the grand-frame - the equivalent of the central bank. This is the puzzle.

\section{Discussion}
\label{sec:discussion}

This extended paradox is a quantum effect, and relies on interference.  At every step, the exchange of momentum between a frame and the system it prepared, or between the grand-frame and each of the subsequent frames, is identical to the classical (coin lending) model we have described. Yet, the overall final result is different. Had all the information been conveyed by angular momentum, the situation would have been the same as in the classical coin situation.  
Classically to simulate our results we would need supplementary information involving hidden variables and overall conspiracy between the hidden variable, the coins and the dice.
Obviously, supplementary information needs to be transferred in the quantum case as well. Quantumly this supplementary information and the ``grand conspiracy'' come via phases when written in the angular momentum representation. Since the phases affect only the conjugate variable to angular momentum, i.e. the angular variables, the supplementary information flows via the angles. In this sense, the supplementary information that propagates is not classical but {\it quantum}.

The key is the intricate interplay between angle and angular momentum which is a hallmark of quantum mechanics. Every interaction which allows for angular momentum exchange must depend upon the relative angles (in the Hamiltonian). However what we have here is deeper than that: we see this interplay already from the beginning.  We have started by discussing an issue concerning angles, namely preparing a state relative to a quantum frame.  This was a question a priori completely disconnected from the issue of angular momentum conservation in individual cases. Yet it turned out that conservation in individual cases is a straightforward consequence of the frame of reference type interaction. Dually, if we would have asked for a momentum conserving interaction, that conserves momentum in individual cases, that is equivalent with preparing a system relative to a frame of reference in angular space.

Another way of looking at the situation is to note that the frames and system preparations of one relative to the other have produced a multi-partite entangled state. Viewed in the angular momentum basis, it is the well defined phases in between the different terms that differentiate this global entangled state from a mixture of various angular momentum terms, which would behave as in our classical model. These phases are due to the conjugate variables, the angles. 

Despite the above, however, the fact that in the branching scenario the angular variables and the entanglement play a role is extremely surprising.  Indeed, they play no role as far as momentum conservation at every individual junction is concerned, as we can see when following the momentum conservation at each junction. Whenever we prepare a particle in superposition of angular momentum, we actually prepare an entangled quantum state between the particle and the preparation device.  When we measure the particle, we find a definite value, and our detailed conservation law shows that the state of the preparation device shifts down by exactly that value, giving us conservation in individual cases.   Everything can be understood in the angular momentum representation, and the phases in the superposition (which encode angular information) are irrelevant. Entanglement, and information encoded in the phases were always present, but they become relevant only when two branches eventually interact.  This is what makes the effect we present here so intriguing and qualitatively different from what happens in a simple chain of reference frames.

\section{Conclusion}
\label{sec:conclusion}

In the present paper we went beyond the usual analysis of quantum frames of reference, and looked at {\it networks} of quantum frames of reference, which are more representative for what happens in Nature, and in real life experiments.  We have analysed the conservation laws in such situations, and we have found qualitatively new effects vs. those exhibited in similar situations such as one frame and one system or chains of frames of reference.  We have also described coordinate transformations that aid in understanding frames of reference, which we believe will become an important tool in this area, in a similar way to how centre of mass/relative coordinates have been very useful in other situations.

We have shown that the recent conjecture, that in quantum mechanics conservation laws hold not only statistically but also in individual cases, is valid not only in simple linear chains of frames of reference, but also in networks of quantum frames of reference.  We found that it works differently from how it did in the linear chain. In a linear chain the detailed conservation law only needs to take into account a particle and its immediate frame of reference: no other frames before the last one needed to be included to ensure conservation. This is illustrated in Fig.~\ref{fig:net-conservation}(a). The detailed conservation law is in this sense local. This is also true in each branch of a network, as in Fig.~\ref{fig:net-conservation}(b). However, in a network of frames which prepares particles which later interact, the detailed conservation law requires us to include all frames in the network from the particles up to and including their first common frame, as in Fig.~\ref{fig:net-conservation}(c). Further, we showed that the flow of information about the conserved quantity is not only through the implied distribution of the conserved quantity, but also through the conjugate variable, the angle, which the frame of reference is used to define.

Still, the way this works is almost miraculous. We can start with any state for the first common frame, any state for the frames in between, any state for the particles, hence there is no fine tuning of any sort between their angles. Yet they always conspire to give conservation in individual cases even in more complex situations such as the branching frames.  

This demonstrates the power of extending the usual statistical conservation law to individual cases, and, arguably, points towards a basic feature of Nature.

\acknowledgments{Daniel Collins, Carolina Moreira Ferrera, Ismael L. Paiva and Sandu Popescu are supported by the European Research Council Advanced Grant FLQuant. Ismael L. Paiva acknowledges financial support from CNPq through the Conhecimento Brasil Program, Grant No. 447051/2024-5.}

\appendix

\section{Example of Bad Coordinate Change}
\label{AppendixBadCoordinates}

In Section~\ref{sec:changeVariables} we said that one has to be more careful when changing variables for angles and angular momentum than we do for linear position and momentum, since angles should be defined between $0$ and $2\pi$ and angular momentum can only have integer quantities.  We shall give an explicit example of how a coordinate change which initially looks reasonable does not work after all.  

For linear position and momentum, it is common to use centre of mass/relative coordinates defined by (taking the masses of F and S to be identical for simplicity):
\beq
\begin{alignedat}{3}
\hx_1 &= \frac{\hx_S + \hx_F}{2} \; &&; \; \hp_1 &&= \hp_F + \hp_S \\
\hx_2 &= \hx_S - \hx_F \; &&; \; \hp_2 &&= \frac{\hp_S - \hp_F}{2},
\end{alignedat}
\eeq
where $\hx$ is linear position and $\hp$ is linear momentum.
Suppose we try naively to do the same for angles and angular momentum.  We define new coordinates by:
\beq
\begin{alignedat}{3}
\htheta_1 &= \frac{\htheta_F + \htheta_S}{2} \; &&; \; \hL_1 &&= \hL_F + \hL_S \\
\htheta_2 &= \htheta_S - \htheta_F \; &&; \; \hL_2 &&= \frac{\hL_S - \hL_F}{2}.
\end{alignedat}
\eeq
One problem with these coordinates is that $\hL_2$ is not always an integer, hence it is not a valid orbital angular momentum: for example if $\hL_S = 1$ and $\hL_P = 0$, then $\hL_2 = 1/2$.  Another problem is that this mapping is not one to one.  For example, suppose we begin with $(\htheta_S,\htheta_F) = (0,\pi)$.  This maps to $(\htheta_1,\htheta_2) = (\pi/2,\pi)$.  However, $(\htheta_S,\htheta_F) = (\pi,0)$ also maps to the same point, $(\htheta_1,\htheta_2) = (\pi/2,\pi)$.  Therefore these coordinates do not work correctly.

One might try to fix this by removing the divide by $2$ from the definition of $\hL_2$, giving
\beq
\begin{alignedat}{3}
\htheta_1 &= \frac{\htheta_F + \htheta_S}{2} \; &&; \; \hL_1 &&= \hL_F + \hL_S \\
\htheta_2 &= \frac{\htheta_S - \htheta_F}{2} \; &&; \; \hL_2 &&= \hL_S - \hL_F.
\end{alignedat}
\eeq
However, in this situation we cannot pick any values $\hL_1$ and $\hL_2$ independently, as e.g. if we pick $\hL_1 = 1$ and $\hL_2 = 0$, that implies $\hL_S = 1/2$, which is no longer an integer.  Thus we cannot write the Hilbert spaces for our new coordinates as the tensor product $\mathcal{H}_1 \otimes \mathcal{H}_2$, and so cannot treat $1$ and $2$ as separate particles.  These coordinates therefore do not work as we would like either.  There simply are no centre of mass/relative coordinates for angles and angular momentum which are equivalent to the centre of mass/relative coordinates for linear position and momentum.

\section{Fourier Transform}
\label{AppendixFourierTransform}

Here we shall demonstrate that our prepared state, whose wavefunction in the angular representation is 
\beq
\psi(\theta_f) \chi(\theta_s - \theta_f),
\eeq
transforms into the wavefunction
\beq
\tilde{\psi}(l_f + l_s) \tilde{\chi}(l_s)
\eeq
in the angular momentum representation, as stated without proof in Eq.~\eqref{coordTransforms} at Section~\ref{sec:changeVariables}.

To do this we simply calculate the Fourier transform of the state,
\beq
\begin{split}
& \tilde{\xi}(l_f,l_s) \\
&= \iint \frac{e^{-i l_f \theta_f}}{\sqrt{2\pi}} \frac{e^{-i l_s \theta_s}}{\sqrt{2\pi}} \psi(\theta_f) \chi(\theta_s - \theta_f) d\theta_f d\theta_s \\
&= \iint \frac{e^{-i (l_f + l_s) \theta_f}}{\sqrt{2\pi}} \frac{e^{-i l_s (\theta_s - \theta_f)}}{\sqrt{2\pi}} \psi(\theta_f) \chi(\theta_s - \theta_f) d\theta_f d\theta_s \\
&= \tilde{\psi}(l_f + l_s) \tilde{\chi}(l_s),
\end{split}
\eeq
where $\tilde{\psi}(l)$ is the Fourier transform of $\psi(\theta)$, 
\beq
\tilde{\psi}(l) = \frac{1}{\sqrt{2\pi}} \int e^{-i l \theta} \psi(\theta) d \theta, 
\eeq
and similarly $\tilde{\chi}(l)$ is the Fourier transform of $\chi(\theta)$.

\bibliography{FrameNetworks}

\end{document}